\documentstyle[]{article}

\begin{document}

\title{On the charged boson gas model as a theory of high Tc superconductivity
\thanks{Alternative title: A model of unconventional superconductivity based on a hypothesis of bound electrons}}
\author{Yanghyun Byun\thanks{Electronic mail: yhbyun@hanyang.ac.kr}\\{\small \it Department of Mathematics, Hanyang University,
 Seoul 133-791, Korea}}

\date{}
\maketitle

\begin{abstract}
We consider a hypothetical bound state of two electrons
even if we leave the binding mechanism unknown.
The bound electron pairs should behave as bosons
and we conclude that the paired electrons are `apparently' in the highest energy states
of all the occupied states.
We call this multi-particle state  an {\it apparent} Fermi surface.
Then the solid in concern must be a type 2 superconductor for a charged boson fluid is theoretically
known to be as such.
Moreover the Bose-Einstein condensation
is known to be closely related to the critical temperatures of high Tc superconductors,
which is consistent with our model.
Most notably the model may describe a superconductor either with or without a Fermi surface.
We propose two experimental methods to detect a bound electron pair,
low energy electron-electron scattering
and photoemission from high Tc superconductors.

\

\noindent {\it Keywords}: bound electron pair,  apparent Fermi surface,
photoemission from HTS, low energy electron-electron scattering

\noindent {\it PACS}: 74.20.-z, 74.20.Mn

\end{abstract}

\tableofcontents

\section{Introduction}
The ideal charged boson gas was considered as a natural model
for superconductivity by Schafroth  \cite{Schafroth} for the first time
in 1955. However the model has become something only of theoretical interest
by the emergence of BCS theory \cite{Bardeen} shortly after.
The original conclusion of Schafroth was that the charged boson gas
should be a superconductor of Type 1.
However a correction was made in  \cite{Friedberg1} by Friedberg  et~al.
to conclude that the model must show superconductivity of Type 2.
The works \cite{Micnas1, Micnas2} by Micnas {\it et~al.}\
also asserted Type 2 superconductivity for a system
of tightly bound electron pairs.
Since high Tc superconductors (HTS's) are of Type 2, these works show that
the charged boson gas
is more relevant to HTS's rather than to conventional ones.

On the other hand the Bose-Einstein condensation (BEC) temperature was the natural candidate
for the critical temperature (Tc)
in the charged boson fluid (CBF) model.
Nevertheless it was many orders of magnitude higher than the Tc's of conventional superconductors
when calculated assuming a sizable fraction of carrier electrons were paired(cf.~p.~xii,~\cite{Alexandrov} or \cite{Rabinowitz}).
This drawback of CBF model is much less serious in the case of HTS's
since they have rather small
densities of paired electrons.
In addition there is the Uemura relation \cite{Uemura1} which asserts that for underdoped cuprates
the Tc's are proportional to $(n_s/m^*)(T\rightarrow 0)$, where $n_s$
is the superfluid density, $m^*$, the effective electron mass and $T$ is the temperature.
This relation has been regarded by some
as implying that the Tc of an HTS is closely related to the BEC of real-space pairs.
In fact Uemura himself, based on the observation
that the 3-dimensional BEC temperatures
are only 4-5 times greater than the Tc's in case of underdoped cuprates,
predicted further that
the Tc's can be properly understood in terms of BEC when
the two dimensional aspect is taken into account together with some other effects (\cite{Uemura2}).

One may regard the partial successes
represented by \cite{Friedberg1,Micnas1,Micnas2} and  \cite{Uemura1}
as an indication that the CBF model
may be a rough approximation to some future successful theory.
Nevertheless we will discuss in this paper  a model
based on tightly bound electrons.
A hypothesis of bound electron pairs is provided in \S 2.1 below.
It leads us to unconventional superconductors which share some well-known
properties with HTS's such as Type 2 superconductivity
and the close relation of the BEC to the Tc's as mentioned in the above (see \S 2.2.3 for more such properties).
Moreover the model appears able to explain the fact that there are HTS's with somewhat
obscure Fermi surfaces
while there are others with undoubtable ones (see \S 2.2, 3 and \S 3.1, 2 below).

Unfortunately the hypothesis does not appear to be understandable
within the known first principles.
It has been long since the known first principles became a reliable system that can be used
to understand all the phenomena in the physical world, with the possible exceptions of the dark matter
and the dark energy. Indeed it is rational to expect
that it is a deep understanding of quantum mechanics of
a many-particle system what is needed to understand the HTS.
Nevertheless this paper does not follow the traditional approach.
Even if most physicists will not agree,
the author thinks that
the current state of affair
might imply that a new first principle is at work in the high Tc superconductivity.

The bound electron pair
can be very elusive (see \S 2.1 and \S 5 below).
However this elusiveness is another ingredient
that allows us to give a thought to such a exotic idea;
otherwise it should have been already known through
the scrutiny by modern physics.
On the other hand the hypothesis is stated only minimally and qualitatively.
It follows that,
even if the hypothesis happens to be real and may describe
correctly some properties of HTS's,
it may not explain many other
experimental facts in details.

Our model of superconductivity has been introduced in \S 2, which is the central
part of the paper.
Then we discuss in \S 3.1, 2 the notable aspect
of the model that it allows a superconductor either with or without a Fermi surface.
The energy of the bound electron pairs
in the free space minus that of two free electrons is referred
to as the excess energy in \S 2.
This energy
seems to be less than $32\, \rm eV $ according to the estimation given in \S 4.
Therefore an experimental support of the bound electron pair
may be obtained in an electron-electron scattering
below $32\, \rm eV $ if there is a resonance as discussed in \S 5.1, 2 below.
This scattering will work only if the excess energy
is a positive value not too small.
We also discuss the possibility that it may not be positive
in \S 2.1 and \S 5.3.
The photoemission from HTS's has been discussed as another method to detect the bound electron pair
in \S 5.3. It may work even if the cross section of bound electron pair formation
is too small for the scattering
or even when the excess energy is negative.
Finally we provide a summary and an outlook of the paper in \S 6.

\section{Superconductivity by bound electrons}
In this section we construct a model for an unconventional superconductor
exploiting a hypothesis that there is a bound state of electrons.
Even if the hypothesis is stated minimally and only qualitatively,
still we may provide
a condition under which the bound state may persist stably.
We will further observe that the bound electrons may accumulate
in a very thin band to form the so-called {\it apparent Fermi surface}.
The apparent Fermi surface appears to consist of `apparently'
highest occupied electron states and therefore the nomenclature is justified.
Our model of an unconventional superconductor is explained in \S 2.2 below exploiting
the arguments of preceding subsections.
The model allows superconductivity either without or with a Fermi surface.
 In \S 2.3 below we discuss a few more issues
regarding the model.

\subsection{The hypothesis}
We state the hypothesis of bound state of electrons as follows:
\begin{quote}
 {\bf Hypothesis B.} There is a bound state of two electrons in the vacuum which is
short-lived and has a size comparable to that of the electron pairs in an HTS.
\end{quote}
To be short-lived in the free space, the bound state should have negative binding energy which we omitted in the above
to avoid redundancy.
That the bound system has a finite size implies that it has an intrinsic structure
which should be taken into account when one considers its interaction with the lattice or with any other system at short distance.

The hypothesis implies a constant:
\begin{quote}
$E_e$ denotes the excess energy of the bound system
relative to two free electrons.
\end{quote}
In fact there might be more than one bound state of two electrons if one ever exists
(see \S 5.2 below). However $E_e$ in the above refers to the lowest one.
Note that the negative binding energy
means $E_e>0$.

Hypothesis~B is far-fetched indeed.
We will not attempt to provide the microscopic mechanism
behind the binding. In fact we do not expect that
the mechanism can be understood exploiting only the known first principles.
Such interactions as originate from polaron, exciton and/or spin fluctuation etc.,
which depend on the lattice and/or the itinerant electrons,
are irrelevant to a binding in the vacuum.
The essence of our hypothesis is
to allow a not-yet-known first principle
to be at work between electrons.

On the other hand we note
that the bound state
has properties, which are necessary, even if not sufficient,
conditions for it not to have been noticed easily.
For instance if the lifetime is short enough
the process of its formation and decay
cannot be easily
distinguished from the usual scattering of two electrons.
In addition we will argue in \S 4 below that the excess energy of the bound state should be
less than $32\, {\rm eV}$.
This energy scale is small enough that the bound state
could have not been noticed
in the myriad of high energy electron-electron scattering
by a resonance.

Now we also consider the possibility $E_e \leq 0$.
In fact the main arguments of the paper, which consist the rest of \S 2 and \S 3,
will be unharmed by the assumption $E_e \leq 0$.
The main reason we assumed $E_e>0$ in the above was that otherwise the electron pair should be stable
and the stability might have made the pair detectable more easily.
In fact even when $E_e$ is $0$ or a small enough positive value,
the life time of the pair will be
rather long since the pair is in effect surrounded by an electric potential wall.
By assuming $E_e$ is negative or a very small positive value
we are assuming that the pair is stable in the free space.
Still it is not easy for the author to imagine a situation
in which the pair can be detected accidentally: First of all the mass-charge ratio
is identical with free electrons and therefore the pair will behave in the same way
as a free electron
both in an electric field and in a magnetic field.
Secondly if an environment, which is other than inside of some solid,
may bring two electrons close enough to allow them to form
a bound pair in spite of the electric repulsion, it must be hot enough
to destroy the pair as well.
Thirdly it will not be easy for an electron pair with a sufficiently large kinetic energy to survive contact
with a condensed matter environment.
Thus one may not expect that the bound electron pairs will be flying around the universe or in the atmosphere of
the earth in a large quantity.
The only modern experimental arrangement in which many free electron pairs might be being produced is photoemission from HTS including ARPES.
However ARPES is designed to detect the fastest electrons while an electron pair is of half their speed when it is fastest.
Another reason why we assumed $E_e > 0$ was
that otherwise HTS's might be more common than
they are in reality, which is not necessarily the case as observed
in the 2nd paragraph from the bottom of \S 5.3 below.

\subsection{The superconductor}
\subsubsection{The stability condition}
To claim any relevance of Hypothesis~B to superconductivity, we need to see first of all
how the bound state may be stable in a solid even when it is short-lived in the vacuum.

We define the energy values $E_0<0$ and $E_t$ as follows:

\begin{quote}
Assume the absolute zero temperature. Then

$E_0$ denotes 2 times the energy
of the lowest unoccupied electron state in the solid and

$E_t$ means
the total energy of the lowest state of a bound electron pair to which $E_e$ is included.
\end{quote}

To make it clearer what we mean by $E_t$, let us note that $E_0$ is the minus of the energy
needed to bring two electrons in the lowest unoccupied state into the free space. Similarly
$E_t$ is the minus of the energy needed to bring the electron pair in its lowest state to two free electrons in the vacuum.
Note that the energy $E_e$  is intrinsic and contributes to the process (assuming $E_e >0$).

An electron which consists a bound pair is not in the competition
 for the allowed states with other unbound electrons.
Here the competition originates
from the Pauli exclusion principle.
Therefore it is possible in some solids that the state of the electron
pair is low enough so that $E_t$ is lower than $E_0$.
If the inequality $E_t <E_0$ holds
and the temperature $T$ is low enough so that the inequality $kT << E_0-E_t$ may hold, then
the bound 2-electron system
will be stable in the solid.
That is, if the bound electron pair
which is in $E_t$ energy state disintegrates, the two electrons should occupy states
whose energy is greater than or equal to $\frac{1}{2}E_0$.
This is not allowed since the energy of the two electron system should increase at least by $E_0 -E_t >0$.
This can be compared to the mechanism which make a neutron stable in a nucleus
while it is unstable in the free space.
Thus we conclude that
\begin{quote}
 If the inequality $E_t<E_0$ holds and the temperature is low enough
 then the bound electron pair exists stably in the solid.
\end{quote}

\subsubsection{The location of $\frac{1}{2}E_t$ in the band structure}
The size of pairing in an HTS is known to be about $1$-$3$ nm.
This size corresponds to a few atoms
or is comparable to one unit cell of a moderately complex lattice.
If the size were zero or can be regarded as zero in the atomic scale,
then its lowest state
would have been the lowest state in the most massive atom in the solid
as a point particle with twice the charge
and with twice the mass of an electron.
If this happens the atom would not be the same as one would expect
from its atomic number.
Indeed bound electrons with zero size would have made
the known atomic phenomena impossible assuming $E_e$ is small enough.
Thus the assumption of finite size is absolutely necessary and it also
allows the bound electrons
any chance to be mobile. We assume in this paper that they are mobile.

Since the electron pairs are charged, any two of them cannot be in the exactly same quantum
state; the electric repulsion will not allow two pairs at the same position.
However we may expect the pair exchange symmetry for the system of the bosons, that is, of the electron pairs.
On the other hand we expect the lattice translation symmetry will remain valid even if only approximately
even in some doped materials.
The translation symmetry implies that degenerate states can be obtained
by applying lattice translations to a single particle state.
Then at the absolute zero temperature and at low enough density
the multi-particle system of the bosons
is in a state which is the tensor product of the degenerate states
(with the symmetrization of the tensor being understood).
Or we may simply take it as an assumption as in \cite{Ioffe} that the mobile bosons
form a thin band.
This homogeneity of energy levels or the thinness of the band
allows us to keep using the same value $\frac{1}{2} E_t$
to represent the apparent energy of the band.

Consider a solid at the absolute zero temperature and assume the inequality $E_t <E_0$ holds.
If there are electrons in states with energies above $\frac{1}{2} E_t$,
then it is energetically favorable for them to form pairwise a bound pair in a state with energy $E_t$.
This means the following:
\begin{quote}
There can be no itinerant electron in a state with energy strictly above $\frac{1}{2} E_t$ and therefore
the bound electrons appear
to form the thin band with the highest energy $\frac{1}{2} E_t$.
\end{quote}

In what follows a band means a continuum of electron states
regardless of whether occupied or not
and regardless of its atomic origin.
This usage of the term appears widely applicable in spite of the complexity of band structure
revealed for instance by the breakdown of conventional band theory
in such systems as Mott insulators  (\cite{Anisimov}).

Now assume that $\frac{1}{2} E_t$ is the same as the energy of an electron state in a partially filled band.
Since the band is partially filled there are electrons with energies infinitesimally close
to $\frac{1}{2} E_0$.
If the inequality $E_t <E_0$ held, all of those electrons
would have been bound pairwise and have fallen into a state with apparent energy $\frac{1}{2} E_t$.
Thus the strict inequality is impossible and
we must have that $E_t =E_0$, assuming the bound electrons exist.
In this case the bound electrons are not stable but in an equilibrium
with the itinerant electrons in states with energy near $\frac{1}{2} E_0$.
Also note that in this case there is a Fermi surface with the Fermi level $\frac{1}{2} E_0=\frac{1}{2} E_t$.

The inequality $E_t < E_0$ may hold only if the following two conditions are satisfied:
(1)~All the bands which contain states with energies lower than $\frac{1}{2} E_t$ are filled.
(2)~All the bands which contain states with energies higher than $\frac{1}{2} E_t$ are unoccupied.
In particular there should be an energy gap between the outermost filled band and the empty band next to it
as it is the case
between the outermost split bands
of a Mott insulator.
Note that the inequality $E_t <E_0$ may hold even when there is no bound electron.
For the bound electrons to exist there should have been some electrons in the band above $\frac{1}{2} E_t$ if it were not
for the bound state of electrons. The band above $\frac{1}{2} E_t$
has become empty because the electrons in it
have bound pairwise to be in the apparent energy state $\frac{1}{2} E_t$.
In particular only in this case the bound electrons may exist stably in the solid.

The discussion so far has led us to the following conclusion, in which
we assume the absolute zero temperature:
\begin{quote}
{\bf Condition S.} If $E_t \leq E_0$, the bound electron pairs may exist in the solid.
If exist, they appear to be concentrated in the highest states with the energy
$\frac{1}{2} E_t$.
\end{quote}

Condition~S above can be divided further into two conditions as follows.

\begin{quote}
{\bf Condition S1.} $E_t = E_0$  if and only if the bound electrons are in an equilibrium with the itinerant electrons.
The equality is also equivalent to the condition that there is a Fermi surface
together with a thin band which has apparently the same energy level.

\noindent {\bf Condition S2.} $E_t <E_0$ if and only if the bound electron pair is stable.
The inequality is equivalent to the conditions that
  $\frac{1}{2} E_t$ lies in the energy gap of the band structure below which all states are occupied
  and above which no state is occupied and that there is no Fermi surface.
\end{quote}

We note again that Condition S2 above does not assert the existence of bound electrons.
For them to exist under the condition, the upper bands should provide the electrons to form the bound pairs.
The upper bands should become empty by doing so.

\subsubsection{The apparent Fermi surface: a model for superconductivity}
If the inequality $E_t \leq E_0$ holds and
the bound electrons exist and are mobile, it seems appropriate for us
to say that the bound electrons have formed an {\it apparent} Fermi surface.
In particular,
as observed in the 3rd paragraph of \S 2.2.2 above,
the bound electrons appear to have accumulated in states with the highest energy
of the occupied electron states.
If the apparent Fermi surface exists in a solid,
it is expected to be a Type 2 superconductor (\cite{Friedberg1,Micnas1,Micnas2}).

In particular, under the equality $E_t = E_0$ the bound pairs are
not stable but in an equilibrium with the itinerant electrons as stated in Condition~S1
above. Thus the system in fact cannot be approximated comfortably
to the CBF.
Both \cite{{Friedberg2}, {Friedberg3}} by Friedberg et.\ al.\
deal with the superconductivity which arises
when the bound state of two electrons is tight and has a random finite life time.
Since the equality $E_t = E_0$  also implies that the bound pairs last only for random finite time intervals,
the works \cite{{Friedberg2}, {Friedberg3}} are applicable.
In particular the work \cite{Friedberg3} shows that the $T_c$
of such a system can be much higher than that of the BCS theory.

Meissner effect of the CBF system has been discussed in \S VI,~\cite{{Friedberg2}}.
The density of pairs appears closely related to the transition temperature
and it has been discussed in \cite{Uemura1} and also in \S I.{\bf C},~\cite{{Friedberg3}}, which implies that
the CBF model
might be compatible with the Tc's of real HTS's.

Furthermore one may assume that the so-called lattice bosons (cf.~\cite{Auerbach1, Auerbach2}) in cuprates
are in fact the bound electron pairs
which are in states tied to the lattice units.
Then more properties
of cuprates may be understood theoretically which include the linear dependence of resistivity
on the temperature (\cite{Auerbach1}) and the abrupt sign change of the Hall conductance near the
critical doping (\cite{Auerbach2}).

\subsection{Further implications of the model}
\subsubsection{Dependence of $E_t$ on the density of bound electrons}
The electrons in states above $\frac{1}{2} E_t$
should pairwise bind and fall into a state with apparent energy $\frac{1}{2} E_t$.
Therefore the density of bound electrons
can be unrealistically high if $\frac{1}{2} E_t$ happens to be a low value.
However it is reasonable to assume
that $E_t$ depends not only on the lattice structure
but also on the density of the bound electrons themselves.
That is,
as the density rises, $E_t$ also rises.
This makes an even better sense when we consider the fact
that in real HTS's
the size and the density of the pair
together imply that
there are overlaps among the pairs that cannot be ignored (cf.~\S 2,~\cite{Uemura2}).
By assuming $E_t$ rises as the density increases
an unrealistically high density of bound electrons can be prevented
and the model can be made compatible
with the known small values of densities of electron pairs
in real HTS's.

\subsubsection{Existence of electrons pairs above $T_c$}
The formation of bound electron pairs is not
directly related to lowering the temperature.
It may be the case that, once the condition $E_t \leq E_0$ is met at the zero temperature,
then the condition may persist in a form corrected  by some thermostatistics
at any temperature
as long as the lattice structure is intact.

Some physicists think
that the electron pair may exist up to $T^*$,
the temperature at which the pseudogap begins to appear (\cite{Yang})
or up to some other temperature $T_{\rm pair}$ such that $T_c < T_{\rm pair} < T^*$ (\cite{Kondo}).
Since the measured $T^*$'s are below $300\, {\rm K}$, this may set the upper limit.
However we note that the origin of the pseudogap is not a settled issue (cf.~\cite{Hinkov}).

\subsubsection{Universality}
The known HTS's are quite diverse and often
in stark contrast in their physical properties.
For instance the overdoped cuprates have fully developed Fermi
surface while the underdoped ones have no Fermi surface or at
best far less conspicuous one. The parent compound of cuprates is
a Mott insulator while it is a metal for iron pnictides.
However Condition~S in the above is not specifically tied
to any of these properties. Therefore
the possibility is open to our model that it may explain all the
diverse HTS's even if it is only under the seemingly unlikely premise
that Hypothesis~B turns out real.

\section{Fermi surface}
In \S 2 above superconductivity  originates from the existence of the apparent
Fermi surface while the real Fermi surface is optional. In this section
we argue that both of  the options may
correspond to some real HTS's.

\subsection{The question of Fermi liquid in underdoped cuprates}
Existence of Fermi surface, and therefore that of Fermi liquid,
in the underdoped cuprates is being supported by quantum oscillation \cite{{Doiron-Leyraud},{Barisic},{Sebastian}}.
One should note however that it is only under the magnetic field $H >H_{irr}$, where
$H_{irr}$ denotes the irreversibility field.

At zero magnetic field Fermi arcs are known to exist by ARPES in the underdoped regime
which are thought to be fragmented Fermi surfaces at the temperature
range $T_c < T < T^*$.
Since the Fermi surface of a two dimensional Fermi liquid
should form a closed loop in the momentum space
there has been a debate regarding the origin.
We note that there is a study \cite{Reber}
which concludes that the Fermi arcs are in fact not related to true Fermi liquids.
Even if the Fermi arc forms a closed loop in some Bi-based cuprate superconductors
at some specific
doping levels which belong to the underdoped regime (\cite{Meng}),
there is a study \cite{King} which show that the loop does
not necessarily imply a Fermi liquid. After all one may say that
there is no decisive evidence for the existence of
Fermi surface under zero magnetic field
in the underdoped regime (\cite{Laforge}).

Thus considering the works \cite{Reber,King,Laforge} it is possible that
superconductivity without the Fermi surface might have been realized in underdoped cuprates.
Note that this corresponds to
the inequality $E_t<E_0$ in our model (Condition~S2 in \S 2.2.2 above).
However to justify this correspondence further we must be able to explain,
starting from the bound electron pair hypothesis, both the Fermi arc and
the result of quantum oscillation.
The latter of course raises a very serious doubt on the validity
of our model. However it might be relevant to the unexpected fact
that a uniform magnetic field necessarily
destroys lattice translational symmetry (p.5, \cite{Auerbach2}).

\subsection{HTS's with Fermi surfaces}
In our model an unconventional superconductor must have an apparent Fermi surface.
A usual Fermi surface is optional; it may or may not exist in an unconventional superconductor.
If it exists,
the Fermi level should be equal
to the apparent level $\frac{1}{2}E_t$
of the apparent Fermi surface. That is, Condition~S1 in \S 2.2.2 above should apply.
Note that the apparent Fermi surface is the one that is responsible
for the superconductivity.
If we assume our model represents HTS's,
this explains the observation of Fermi surface below the $T_c$ for the HTS's with
Fermi surface above the $T_c$ (cf.~\cite{Chang,Plate}).
Note that the Fermi surface is destroyed
by the emergence of superconductivity
in conventional superconductors.

In fact ARPES might be seeing the apparent Fermi surface since
the bound electrons may constitute a source of the most energetic electrons
in photoemission.
However the anomaly originating from apparent Fermi surface
must be less conspicuous in the presence of usual Fermi surface
than in the case when there is only apparent Fermi surface
as in 3.1 above.
We note that there are studies  such as \cite{Chang,Awanda,Castro} which report some anomalies
in the Fermi surfaces of HTS's.

\section{A rough estimation of the excess energy}
Note that our model of superconductivity in the above can be real only if Hypothesis~B is real.
Furthermore it can be realistic only if the excess energy $E_e$ is small enough as to allow
the inequality $E_t \leq E_0$ in some solids.
Recall that $E_e$ is a universal coefficient in our model by which
the energy of the bound electron pair
in the free space is greater than that of two free electrons (\S 2.1).
In this section we propose an upper bound for $E_e$.
We begin by considering the energies;
\begin{quote}
$E_s$ denotes the energy of the lowest state of the bound electron pair which is the sum of the potential
energy and the center-of-mass kinetic energy.

$E_i$ denote the gain of energy coming from distorting the structure of the bound state
by putting the pair in a lattice.
\end{quote}
Then we have that $E_t=E_e +E_s+E_i$.
Also we assume $E_i>0$.
Then recall Condition~S in \S 2.2.2 which demands
the inequality $E_t \leq E_0$ for an unconventional superconductivity. Therefore we have that $E_e \leq E_0-E_s-E_i$.
Thus  $E_0 -E_s$ is an upper bound for $E_e$.
Note that the upper bound $E_0 -E_s$ is better if the value $E_i>0$  is smaller.
However we do not know how to estimate $E_i$. Thus it is difficult to tell how good the upper bound $E_0-E_s$ is for $E_e$.

We will put $E_0=2ⅹ(-4\, {\rm eV})$, where $4\, {\rm eV}$ is chosen as the typical work function of a metal.
Then the upper bound depends only on the estimation of $E_s$.

\subsection{Basic facts and assumptions for an estimation of $E_s$}
First of all we assume the interaction of the bound electrons with the lattice is electric. Then we observe the following facts:

\begin{quote}
(1)	The bound electron pairs are apparently mobile in HTS's.

(2)	The electrons in the bound state are not subject to the same constraints in their allowed states as the other electrons.

(3)	The inner space of an atom is positively charged and apparently provides potential energy
to a point particle with charge $-e$ (by $e$ we mean the charge of a proton). Let $Z$ denote the atomic number. Let $r$ denote the distance from the nucleus.
Then the potential energy is proportional to $- {e\over r}$ near the outermost region of the atom and to $- {eZ\over r}$ near the nucleus.
\end{quote}

We take the mobility condition in (1) above as meaning that the kinetic energy is zero.
Thus only the potential energy contributes to $E_s$. We need a lower bound of $E_s$
to have an upper bound of $E_e<E_0 -E_s$.

\subsection{An estimation}
For a calculation of the lower bound for $E_s$, we assume as follows:
\begin{quote}
The wave function of the bound electron system is such that the electrons are more or less evenly distributed
throughout the space occupied by the solid regardless of whether it is the inner space of the atoms or the outer space.
\end{quote}

We also consider a solid specified as follows:
\begin{quote}
(1) The lattice structure is simple cubic with edge $\rm 0.3\, nm$.

(2) There is an ion with charge $e$ at each vertex
  and the radius of the ion is $\rm 0.15\, nm$.

(3) In the inner space of the ion the potential of a particle with charge $-e$
at distance $r$ from the nucleus can be approximated by $-\frac{\epsilon}{r^2}$ with
$\epsilon=0.22\, \rm eV\cdot nm^2$.

(4) In the outer space the potential for a particle of charge $-e$
is homogeneously $ -9.6\rm \, eV$.
\end{quote}

We may write $-\frac{\delta}{r}$ for the potential of an electron at distance $r$ from a proton,
where $\delta = \rm 1.22\, eV\cdot nm$.
In fact the constant $\epsilon$  in (3) above is chosen
so that $-\frac{\epsilon}{r^2}=-\frac{\delta}{r}$  when $r=\rm 0.15\, nm$.
If $Z$ is the atomic number of the ion,
the inequality $-\frac{Z\delta}{r} \leq -\frac{\epsilon}{r^2}\leq -\frac{\delta}{r}$ holds
when $\frac{a}{Z} \leq r \leq a$, where $a=\rm 0.15\, nm$.
Thus $-\frac{\epsilon}{r^2}$ is a reasonable choice at least in the interval $\frac{a}{Z} \leq r \leq a$.
Furthermore since our goal is to find a rough lower bound for $E_s$, our choice in (3)
can be justified in the whole interval $0< r \leq a$.
Also note that the homogeneous potential $\rm -9.6\, eV$ for the outer space in (4) makes a good sense:
(i) The equality holds that $-\frac{\epsilon}{a^2}= -9.6\, \rm eV$.
(ii) The free electrons present in the outer space will make the potential nearly homogeneous.
(iii) $9.6\, \rm  eV$ is a value close to the sum of the typical work function $4\, \rm eV$
and the typical Fermi energy $\rm 4\, eV$ of a metal.

Then we obtain $\rm -31\, eV$ as the contribution of the inner space.
By adding the contribution of the outer space we obtain $E_s =\rm -40\, eV$
as a lower bound.
Note that the calculation implies that $E_s$ will be a larger negative value
if the ions are more densely packed. In other words, if the ratio of inner space of ions
to the total volume of the solid is greater, then the calculation will give a larger negative value for $E_s$.

The values, $E_0=-8\, {\rm eV}$ and $E_s= -40\, {\rm eV}$, mean that we have $E_0-E_s=32\, {\rm eV}$.  We conclude that $E_e< 32\, {\rm eV}$.

\subsection{The screening effect}
In fact the potential $-\frac{\epsilon}{r^2}$ with
$\epsilon=0.22\, \rm eV\cdot nm^2$ in (3) of the previous section
cannot be a good approximation.
For instance the deep inner space of an atom with a large atomic number
may not provide such a large energy gain as implied by $-\frac{\epsilon}{r^2}$
to a point particle with charge $-e$. This is because the screening of positive charge
of the nucleus will raise the energy levels of all the outer electrons and some portion
of the energy gain
by nearing the nucleus will be compensated by this screening.
Considering the screening effect, it is not clear and appears not known
to what extent a point particle with charge $-e$, which need not be an electron,
will feel an attractive force toward the nucleus inside an atom.
In any case $32\, {\rm eV}$ based on the potential $-\frac{\epsilon}{r^2}$ appears
overly generous upper bound for $E_e$
even in the hypothetical solid of the previous section.

On the other hand we noted that the calculation of $E_s$ in the above depends on the ratio of inner space
of ions and atoms to the total volume of the solid which is closely related to the atomic number density of the solid.
However the atomic number density does not vary greatly from a solid to another and the solid with the lattice structure described
in the above is realistic enough. Also note that $E_e +E_i < -E_s +E_0$ is the original inequality
from the opening words of the section
and that we expect $E_i>0$.  Thus the upper bound $32\, \rm eV$ for $E_e$
in the above appears quite generous.

\section{Detection of bound electrons}
\subsection{Low energy electron-electron scattering}
Let us consider an electron-electron beam scattering arrangement.
If Hypothesis~B in \S 2.1 above are real, one may expect
that there will be a resonance for the formation of the bound states
when the kinetic energy of each beam is  $\frac{1}{2} E_e$.
Note that we have proposed an upper bound $E_e < 32\, {\rm eV}$ for $E_e$
in \S 4.3 above.
The bound state
will shortly
decay into two free electrons.
We do not know whether or not the decay will
necessarily accompany emission of photons.
In any case the event cannot be easily distinguished from the usual electron-electron scattering,
which means that there is a strong background noise for the scattering experiment.
It is not clear whether the resonance
can be detected in spite of this noise.

One may reduce the noise of usual scattering to some extent
by concentrating on the events such that
two electrons are scattered off from each other in directions perpendicular
to the beams.
This is because electrons are fermions.
In fact there are some graduate texts of quantum mechanics
which explicitly deal with fermion-fermion scattering.
According to them the noise can be reduced by this arrangement
to a quarter of the value when electrons were bosons instead of fermions.

The scattering method seems useful only when the cross section of bound pair formation is large enough and
the life time of bound electrons is short enough.
Otherwise the photoemission from HTS's might be a more appropriate method
as we will discuss in \S 5.3 below.

\subsection{Removing the noise}
In principle the noise in \S 5.1 above can be made to vanish
by taking into account the spin states as well in addition to the momenta.
However this method is useful only
under the following assumptions:

\begin{quote}
(1) A large fraction of bound electrons decay without
emitting any photon.

(2) The possibility is not significantly
suppressed that the two electrons from a decay may be in the same spin state.
\end{quote}

\noindent
In fact the second assumption might appear impossible; the two electrons
in a bound state
are expected to have spin states opposite to each other
as fermions of the same species.
Therefore if both (1, 2) above are satisfied it will
be a surprise by itself and in fact will imply the two electrons in the bound state are
different from each other in some way.
The binding energy of the pair with the same spin is expected to be greater by a small value than the bound pair
with the opposite spin.
The reason why we expect the difference to be small is
that
the size of the bound pair is larger than
the atomic scale by an order.

The vanishing of the noise can be achieved as follows:
Assume we have arranged the two beams so that they are polarized respectively upward
and downward when the $z$-axis is chosen perpendicular to the beams. Then we concentrate
on the events in which the electrons are scattered off elastically
and perpendicularly to the beams.
Furthermore let us choose the beam line as the $x$-axis. Then in addition we consider only the case
when both scattered electrons are in spin up (or down) state with respect to the $x$-axis.
Then the contribution of usual scattering to this event should vanish.
In fact this vanishing will be achieved even when both of the beams are polarized upward (or downward)
with respect to the $x$-axis, which illustrates somewhat dramatically the fact that spin is not conserved
in scattering of identical fermions.

The proof of this vanishing is as follows:
Let $R$  denote the reflection of 3-space with respect to the $yz$-plane
and let ${\rm\bf R}$  be the corresponding quantum
transformation. Let $|p, \pm_z\rangle$ denote free electron states.
Then it is straightforward to see that
$${\rm\bf R}|p, \pm_z\rangle =i |Rp, \mp_z\rangle.$$
Also it is not difficult to see that
$${\rm\bf R}|p,\pm_x\rangle=\pm i|Rp, \pm_x\rangle.$$
Let $p_1, p_2$ represent the initial electron 4-momentums which are related by $p_2 =Rp_1$
and $p_1', p_2'$ be the final electron 4-momentums which satisfies $Rp_i'=p_i'$, $i=1,2$.
Consider a Feynman diagram FD and let $S_{\rm FD}$ denote the scattering operator represented by
FD. Note that
$S_{\rm FD}$ is invariant under $R$ (cf.~p.~76, \cite{Weinberg}).
Now we have:
\begin{eqnarray*}
\lefteqn{\langle p_1', +_x; p_2',+_x| S_{\rm FD} |p_1, +_z; p_2, -_z \rangle}\\
  &&  =\langle p_1', +_x; p_2',+_x|{\rm\bf R}^* S_{\rm FD}{\rm\bf R}|p_1, +_z; p_2, -_z \rangle\\
  &&  =\langle p_1', +_x; p_2',+_x|S_{\rm FD}|p_2, -_z; p_1, +_z \rangle.
\end{eqnarray*}
The last expression is the contribution of the Feynman diagram
obtained by exchanging the initial electrons. Since electrons are fermions
the contributions of the two Feynman
diagrams cancel each other completely. Note that this cancellation should work also
when both of the electrons are initially in $|+_x\rangle$ (or $|-_x\rangle$) spin states.

\subsection{Photoemission from HTS's}
Consider the work function which is $-{{1}\over{2}} E_t$
when there is a Fermi surface.
In general, when only one
of the bound electrons is emitted and the other enters the $\frac{1}{2} E_0$ state,
the energy needed is $-{{1}\over{2}} E_t + {{1}\over{2}} (E_0 -E_t)={{1}\over{2}} E_0 -E_t$, which is slightly larger value than
$-{{1}\over{2}} E_t$.
It follows that, if $E_t < E_0$ and $e_1$ denotes the energy of highest state in the filled
state below $\frac{1}{2}E_t$, then the smaller of $- e_1$ and ${{1}\over{2}} E_0 -E_t$ is the work function.

In any case if the photon energy reaches $E_e -E_t$
an extra channel of photoemission may be at work:
the bound electron pair itself may be emitted and subsequently disintegrate
into two free electrons with additional momentums in opposite directions
corresponding to the kinetic energies ${\frac{1}{2}}E_e$.

Detection of this channel of photoemission is not guaranteed simply by existence
of the apparent Fermi surface. First of all non trivial number of electron pairs should survive
the impact with the photons and be emitted to the vacuum.
Then the appropriate arrangement for the detection will vary mainly
depending on the lifetime of the bound state.
The merit of this method is that it might work even when the cross section
of bound pair formation is too small or the lifetime of the bound pair is too long
for the low energy electron-electron
scattering.

On the other hand we have assumed $E_e >0$ so far throughout the paper. The only exception is the last paragraph of \S 2.1 above
in which we explicitly considered the possibility $E_e \leq 0$.
However we have not in fact exploited the assumption $E_e >0$ at all except in
the above two subsections.
Nevertheless there were two reasons that $E_e>0$ looked more plausible:
The first one was that
if $E_e \leq 0$ the bound pair is stable and therefore it seemed more likely that
it could have been detected accidentally.
This is not necessarily so (see the last paragraph of \S 2.1 above).
The second one is that if $E_e \leq 0$ it might be much easier for a solid to be a superconductor
and unconventional superconductors should have been more common than
they are actually found to be.
However there is $E_i$, the gain of energy coming from distortion of the structure of the bound pair
by putting it in a lattice, which was introduced in the opening words of \S 4 above.
The energy $E_i$ may potentially prevent most of solids from becoming unconventional superconductors.
In case $E_e\leq 0$
the scattering experiment will not work since there cannot be a resonance.

Even if the bound electron pairs are emitted in sufficient quantity
by the photoemission from HTS's and
are stable in the free space,
its detection does not appear necessarily
an easy task.
One may start from the photoemission from an HTS
and then exploit intense homogeneous light beams to isolate the pairs from the free electrons.
A spin measurement can be decisive in the end.

\section{Summary and outlook}
The assumption of bound electron pairs, which may persist even in the vacuum,
with finite size (\S 2.1) leads us to a model of unconventional
superconductors.
In fact under an inequality ($E_t \leq E_0$ in \S 2.2.2) and at zero temperature
the bound electrons accumulate
in a very thin band which we call the `apparent
Fermi surface'. It appears to consist of
an apparently highest states.
For the model to be realistic, the bound electron pair should of course be real.
In addition the excess energy ($E_e$ in \S 2.1) should be small enough for $E_t \leq E_0$
may hold. We estimated that $E_e$ should be much less than $32\, \rm eV$ in \S 4.
The inequality $E_t < E_0$ implies that the unconventional superconductor does not
have a (usual) Fermi surface while the equality $E_t = E_0$ implies that it has one (\S 2.2.2, 3).
The former may correspond to underdoped cuprates while the latter
corresponds to any HTS with a Fermi surface (\S 3.1, 2).
The low energy electron-electron scattering (\S 5.1, 2) seems to be a direct method
to test Hypothesis~B. However the cross section of the formation
of the bound electron pairs could be too small for the scattering to work.
It is also possible that $E_e$ may be too small and even negative, which also make
the scattering impossible.
We may consider the photoemission from HTS's as another way to test the existence
of bound electron pairs in the vacuum.
This method might work regardless of the size of the cross section
and even for small (including negative) values of $E_e$ (\S 5.3).

If an experiment confirms the existence of bound electron pairs in the free space,
it most likely implies
a new first principle.
From the view point
of general physics it will be a cataclysm; it
dismantles the belief that the electron is fully understood within
the framework of the standard model.
On the other hand a theoretical understanding of high Tc superconductivity
will be still only
at a beginning stage.
Some experimental data on HTS's can be exploited to study the intrinsic structure
of the bound electron pair, its interactions with other pairs, with itinerant electrons and most importantly with the lattice. Then one may
attempt to build a detailed theory of unconventional superconductivity by introducing a quantum
theory of a many-body system.

\end{document}